\documentclass[prl,twocolumn,aps,amsmath,amssymb,longbibliography,floatfix]{revtex4-2}

\usepackage{graphicx}
\usepackage{amsmath,amssymb,bm,epsfig}
\usepackage{dcolumn}
\usepackage{mathtools}
\usepackage{esint}
\usepackage{xcolor}

\usepackage{tikz}
\usetikzlibrary{datavisualization}
\usetikzlibrary{datavisualization.formats.functions}
\usetikzlibrary{decorations.pathmorphing,patterns}
\usetikzlibrary{decorations.pathreplacing,calligraphy}
\usetikzlibrary{decorations.markings}
\usetikzlibrary{calc,patterns,angles,quotes}
\usetikzlibrary{shapes.geometric}
\usetikzlibrary{arrows.meta}

\newcommand\identity{1\kern-0.25em\text{l}}
\newcommand\zeroo{0\kern-0.4em\text{0}}

\graphicspath{ {./Figures/} }

\begin{document}

\preprint{arXiv:2503.00562}

\title{Quantum Lamb model}

\author{Dennis P. Clougherty}
\author{Nam H. Dinh}
\affiliation{
Department of Physics\\
University of Vermont\\
Burlington, VT 05405-0125 USA}

\date{June 17, 2025}

\begin{abstract}
H.~Lamb considered the classical dynamics of a vibrating particle embedded in an elastic medium before the development of quantum theory.  Lamb was interested in how the back-action of the elastic waves generated can damp the vibrations of the particle.   We propose a quantum version of Lamb's model.  We show that this model is exactly solvable by using a multimode
Bogoliubov transformation.   We find that the exact system ground state is a multimode squeezed vacuum state, and we obtain the exact Bogoliubov frequencies by numerically solving a nonlinear integral equation.  A closed-form expression for the damping rate of the particle is obtained, and it agrees with the result obtained by perturbation theory.  The model provides a solvable example of the damped quantum harmonic oscillator. 
 \end{abstract}

\maketitle
 
\newpage

\section{Introduction}
  
Advances in the fabrication and characterization of simple mechanical systems in the nanoscopic and mesoscopic regime have facilitated experimental and theoretical investigations \cite{dykman,cleland-book} into some of the foundational principles of quantum mechanics.  Prominent examples of such systems include vibrating beams and mirrored surfaces that interact with laser light through its radiation pressure (optomechanics) \cite{aspelmayer}, mechanical resonators coupled to electronic devices (nanoelectromechanics) \cite{geller,blencowe}, and interacting mechanical resonators (quantum acoustodynamics) \cite{clougherty14,dpc2017,cleland}. In addition to providing a path to explore quantum science and the limits of precision measurement, such systems might be used to fashion new quantum sensors and devices for manipulating quantum information \cite{schwab,mechanicalqubit}.

\begin{figure}[h]
\[\begin{tikzpicture}
\filldraw [pattern=north east lines] (0,-2.5) rectangle (7,-2);
\filldraw [pattern=north east lines] (0,3) rectangle (3,3.5);
\draw [fill=black] (1.5,1.5) circle[radius=3pt] node[xshift=-1mm,left] {\Large{$m$}};
\draw (1.5,0) -- (1.5,0.1);
\draw[decoration={aspect=0.3, segment length=1.5mm, amplitude=2mm,coil},decorate] (1.5,0.1) -- (1.5,1.3) node[midway,xshift=5mm] {\Large{$\kappa_c$}}; 
\draw (1.5,1.3) -- (1.5,1.5);
\draw (1.5,1.5) -- (1.5,1.7);
\draw[decoration={aspect=0.3, segment length=1.25mm, amplitude=1.5mm,coil},decorate] (1.5,1.7) -- (1.5,2.9) node[midway,xshift=4mm] {\Large{$\kappa$}}; 
\draw (1.5,2.9) -- (1.5,3);
\draw (1.2,-2) rectangle (1.25,0.75);
\draw (1.225,0) ellipse (7pt and 3.5pt);
\draw (6.7,-2) rectangle (6.75,0.75);
\draw (6.725,-0.5) circle[radius=1pt] node[xshift=-0.1mm,left]{};
\draw [thick] (6.7,-0.5) cos (1.475,0);
\end{tikzpicture}\]
\caption{\label{fig:setup} Schematic of a generalization of the classical Lamb model. Bead of mass $m$ at $x=0$ is constrained to move in the vertical direction.  The vibrating bead is coupled by a spring to a long string under tension $\tau$. The vibrating bead creates transverse acoustic waves on the string ($\ell\gg c/\omega_0$). The bead subsequently undergoes damped harmonic motion.}
\end{figure}
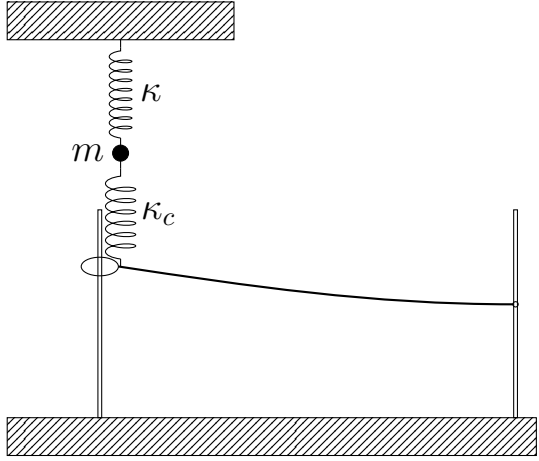

We consider a mechanical system whose first study predates the development of quantum mechanics.  In 1900, Lamb \cite{lamb00} considered the dynamics of a vibrating particle embedded in an elastic medium.  The back-action of the elastic waves generated by the vibrations of the particle work to damp those vibrations creating a damped harmonic oscillator.  In this work, we study a quantum version of Lamb's model and focus on the dynamics of the vibrational decay. Figure~\ref{fig:setup} shows a schematic consisting of a vibrating bead coupled by a spring to a long string under tension that serves as the classical basis of the model.  

There have been other formulations of the damped quantum harmonic oscillator.  Feshbach and Tikochinsky \cite{feshbach} introduced an auxiliary variable into the lagrangian of a harmonic oscillator to get the desired effective equation of motion for the damped oscillator.  They then proceeded by canonical quantization to obtain a quantum description of the damped harmonic oscillator.  The auxiliary variable presumably functions as a single, effective environmental degree of freedom, but the connection to the microscopic physics is not made.  

Caldeira and Leggett \cite{leggett} separate the system into a sum of two subsystems (oscillator and bath) plus an interaction.  Using a path integral description, the bath degrees of freedom can be integrated out to give a general quantum formulation of dissipative systems.  Yurke \cite{yurke} specifically considered a Lamb-type model that is a special case of the model considered here.  (We will recover Yurke's results by allowing the spring that couples the bead motion to the string to be suitably stiff.)  Yurke considered a string with a point mass at one end.  The point mass is also coupled to a spring with a fixed end.  The mass-loaded string then has a time-dependent boundary condition.  As a result, the normal modes are nonorthogonal.  Yurke overcame this by finding an appropriate weighting factor to use in redefining the inner product so that generalized orthogonality can be applied.  He then quantized the model in the standard way.

Following Caldeira and Leggett \cite{leggett}, the model considered in this work expresses the Lamb Hamiltonian as a sum of two subsystems (oscillator and string) plus a coupling term.  Since the coupling is bilinear in operators, the Hamiltonian is exactly diagonalizable with the use of a multimode Bogoliubov transformation.  We find explicit expressions for the coefficients that diagonalize the Hamiltonian.  Using the symplectic properties of the transformation, we confirm that our results satisfy the necessary identities.  We then derive a nonlinear equation whose solution yields the Bogoliubov frequencies, and we use it to numerically calculate the symplectic spectrum of the model.  

We show that the ground state of the quantum Lamb model is a nonclassical state--a multimode squeezed vacuum state-- and we relate this ground state to the uncoupled states (transverse phonons of the string and vibrons of the bead) of the system.  Squeezed states can serve as a quantum resource for precision sensing applications; for example, gravitational wave detection relies on squeezing to perform displacement measurements where uncertainty in momentum is sacrificed in favor of reduced uncertainty in position.  Caldeira and Leggett \cite{leggett} found in their studies of the damped quantum oscillator, the uncertainty in position for the ground state is reduced with increasing damping, strongly so in the overdamped regime.  This result is consistent with a squeezed ground state.

We then study the dynamics of the vibrational decay of the bead.  Dissipation emerges in the thermodynamic limit where the number of string modes $N$ becomes infinitely large ($N\to\infty$).  In this limit, vibrational energy of bead can be radiated away.  We then obtain an explicit expression for the decay rate, and we calculate the spectral distribution of single bogoliubon emission, a product of the bead decay.  We show that for weak coupling strength $g$, the decay rate calculated agrees with both the classical result and the golden rule.

\section{Hamiltonian}

The Hamiltonian of the system in Fig.~\ref{fig:setup} is 
\begin{equation}
H=\sum_{\alpha=0}^N \omega_\alpha a_\alpha^\dagger a_\alpha-\left(a_0+a_0^\dagger\right)\sum_{n=1}^N \gamma_n\left(a_n+a_n^\dagger\right)
\label{ham}
\end{equation}
where $a_n^\dagger$ ($a_n$) creates (annihilates) a transverse acoustic phonon on the string, and $a_0^\dagger$ ($a_0$) creates (annihilates) a vibron on the bead.  (We use index notation where (greek) $\alpha=0, 1, 2, \dots$, while (roman) $n=1, 2, \dots, N$, and work with natural units where $\hbar=1$.) $N$ is the number of vibrational modes for the string.  We are ultimately interested in the limit $N\to\infty$ in order to obtain a description of the damped bead oscillator.

The frequency $\omega_0=\sqrt{(\kappa+\kappa_c)/m}=\sqrt{\omega_b^2+\omega_c^2}$ is the bead vibrational frequency with the string fixed at $x=0$, while $\omega_n$ are the vibrational frequencies of the string (tension $\tau$, length $\ell$, mass density $\sigma$, transverse speed of sound $c$) subject to a spring boundary condition at $x=0$ and a fixed condition at $x=\ell$.

The coupling parameters $\gamma_n$ can be expressed in terms of physical parameters of the model \cite{SM}
\begin{equation}
\gamma_n=\omega_s\sqrt{\frac{\nu}{\omega_0}}\sqrt{\frac{k_n\ell}{(k_n\ell)^2+(k_s\ell)^2}}\frac{1}{\sqrt{1-\frac{k_s\ell}{(k_n\ell)^2+(k_s\ell)^2}}}
\label{gamma}
\end{equation}
where $\omega_s=c k_s=\frac{c \kappa_c }{\tau}$ and $\nu=\frac{\tau}{2mc}$.
The wavenumber $k_n$ is a solution of the transcendental equation $\tan k_n\ell=-\frac{\tau}{\kappa_c}k_n$.

\section{Bogoliubov transformation}

We diagonalize the Hamiltonian in Eq. \ref{ham} with the use of a multimode Bogoliubov transformation.  We look for a linear transformation (and its inverse) with the following form:
\begin{eqnarray}
b_\alpha&=&\sum_\beta\left(M_{\alpha\beta} a_\beta+N_{\alpha\beta} a_\beta^\dagger\right)\\
a_\alpha&=&\sum_\beta\left(U_{\alpha\beta} b_\beta+V_{\alpha\beta} b_\beta^\dagger\right)
\end{eqnarray}
where $b_\alpha^\dagger$ ($b_\alpha$) creates (destroys) a Bogoliubov excitation (bogoliubon) and \textsf{M}, \textsf{N}, \textsf{U}, and \textsf{V} are $(N+1)$-dimensional square matrices whose elements are the coefficients of the transformation.  (The string with length $\ell$ is a system of $N$ discrete atoms.)

We require that the transformation preserve the boson commutation rules  
$\left[b_\alpha, b^\dagger_\beta\right]=\delta_{\alpha\beta}$.  As a result \cite{coherentstates,Adrian2004}, the coefficients can be grouped to form a $2(N+1)$-dimensional symplectic matrix $\textsf{T}\in \text{Sp}(2(N+1),\mathbb{R})$:
\begin{equation}
\textsf{T}\equiv \begin{pmatrix}
\textsf{M} & \textsf{N}\\
\textsf{N} & \textsf{M}
\end{pmatrix}.
\label{bgt}
\end{equation}
$\textsf{T}$ then satisfies the symplectic condition $\textsf{T} \textsf{J} \textsf{T}^T= \textsf{J}$
where the symplectic form $\textsf{J}$ can be represented as
\begin{equation}
\textsf{J}\equiv \begin{pmatrix}
\ \ \zeroo & \identity\\
-\identity &\zeroo
\end{pmatrix}.
\label{form}
\end{equation}
A number of useful coefficient identities follow \cite{SM} from the symplectic structure on this Fock space; for example,  the inverse of the transformation matrix \textsf{T} can be obtained simply from the symplectic condition (together with $\textsf{J}^2=-\identity$):
\begin{eqnarray}
\textsf{T}^{-1}&=&- \textsf{J} \textsf{T}^T \textsf{J}\\
&=&\begin{pmatrix}
\textsf{M}^T & -\textsf{N}^T\\
-\textsf{N}^T & \textsf{M}^T
\end{pmatrix}.
\label{ibgt}
\end{eqnarray}
Hence, we conclude that the coefficients of the inverse transformation satisfy $\textsf{U}=\textsf{M}^T$ and $\textsf{V}=-\textsf{N}^T$.  We summarize the explicit form for the transformation in Table~\ref{table:params}.  The detailed calculation of the coefficients is outlined in the Supplemental Material \cite{SM}.

The following Hamiltonian results 
\begin{equation}
H=\sum_\alpha \Omega_\alpha b_\alpha^\dagger b_\alpha
\label{newham}
\end{equation}
where the Bogoliubov frequencies $\{\Omega_\alpha\}$ satisfy the following summation equation:
\begin{equation}
\Omega_\alpha^2=\omega_0^2+ 4\omega_0\sum_q\frac{\gamma_q^2 \omega_q}{\Omega_\alpha^2-\omega_q^2}
\label{spectrum}
\end{equation}
By using the pole expansion form (Mittag-Leffler) for cotangent,  it is straightforward to show that in the thermodynamic limit ($\ell, N\to\infty$ and $\frac{N}{\ell}\to \frac{\omega_d}{\pi c}$),  Eq.~\ref{spectrum} gives Yurke's transcendental equation \cite{yurke} for the special coupling case of $\omega_c=\omega_c^*\equiv\sqrt{\frac{4}{\pi}\nu\omega_d}$
 \begin{equation}
 \Omega_\alpha^2=\omega_b^2+2\nu\Omega_\alpha\cot\frac{\Omega_\alpha\ell}{c}
 \label{yurke-freq}
 \end{equation}
 The frequency $\omega_d$ is recognized as the Debye frequency of the string (the high-frequency cutoff for the string).

\begin{widetext}
\begin{table}[h]
\centering
\caption{\label{table:params} Coefficients for the Bogoliubov transformation $M_{\alpha\beta}$ and $N_{\alpha\beta}$ ($\alpha, \beta=0, 1, \dots, N$ and $k, q=1, 2, \dots,N$). Coefficients for the inverse transformation can be obtained from the transpose relations $U_{\alpha\beta}=M_{\beta\alpha}$ and $V_{\alpha\beta}=-N_{\beta\alpha}$ (see Supplemental Material \cite{SM}).\\}
\centering
\resizebox{.7\textwidth}{!}{%
\begin{tabular}{|c|c|c|c|} 

\hline
& $(\alpha0)$ & $(\alpha k)$  \\ 
\hline
$M\ $ & $\frac{\Omega_{\alpha}+\omega_0}{\sqrt{4\omega_0\Omega_{\alpha}}}\frac{1}{\sqrt{1+4\omega_0\sum_q\frac{\gamma_q^2\omega_q}{\left(\Omega_\alpha^2-\omega_q^2\right)^2}}}$ & $-\frac{2\omega_0\gamma_{k}}{(\Omega_\alpha-\omega_{k})}\frac{1}{\sqrt{4\omega_0\Omega_\alpha}}\frac{1}{\sqrt{1+4\omega_0\sum_q\frac{\gamma_q^2\omega_q}{\left(\Omega_\alpha^2-\omega_q^2\right)^2}}}$ \\
\hline 
$N\ $ & $\frac{\Omega_{\alpha}-\omega_0}{\sqrt{4\omega_0\Omega_{\alpha}}}\frac{1}{\sqrt{1+4\omega_0\sum_q\frac{\gamma_q^2\omega_q}{\left(\Omega_\alpha^2-\omega_q^2\right)^2}}}$ & $-\frac{2\omega_0\gamma_{k}}{(\Omega_\alpha+\omega_{k})}\frac{1}{\sqrt{4\omega_0\Omega_\alpha}}\frac{1}{\sqrt{1+4\omega_0\sum_q\frac{\gamma_q^2\omega_q}{\left(\Omega_\alpha^2-\omega_q^2\right)^2}}}$  \\
\hline
\end{tabular}%
}
\end{table}
\end{widetext}

\begin{figure}[h]
\includegraphics[width=7cm]{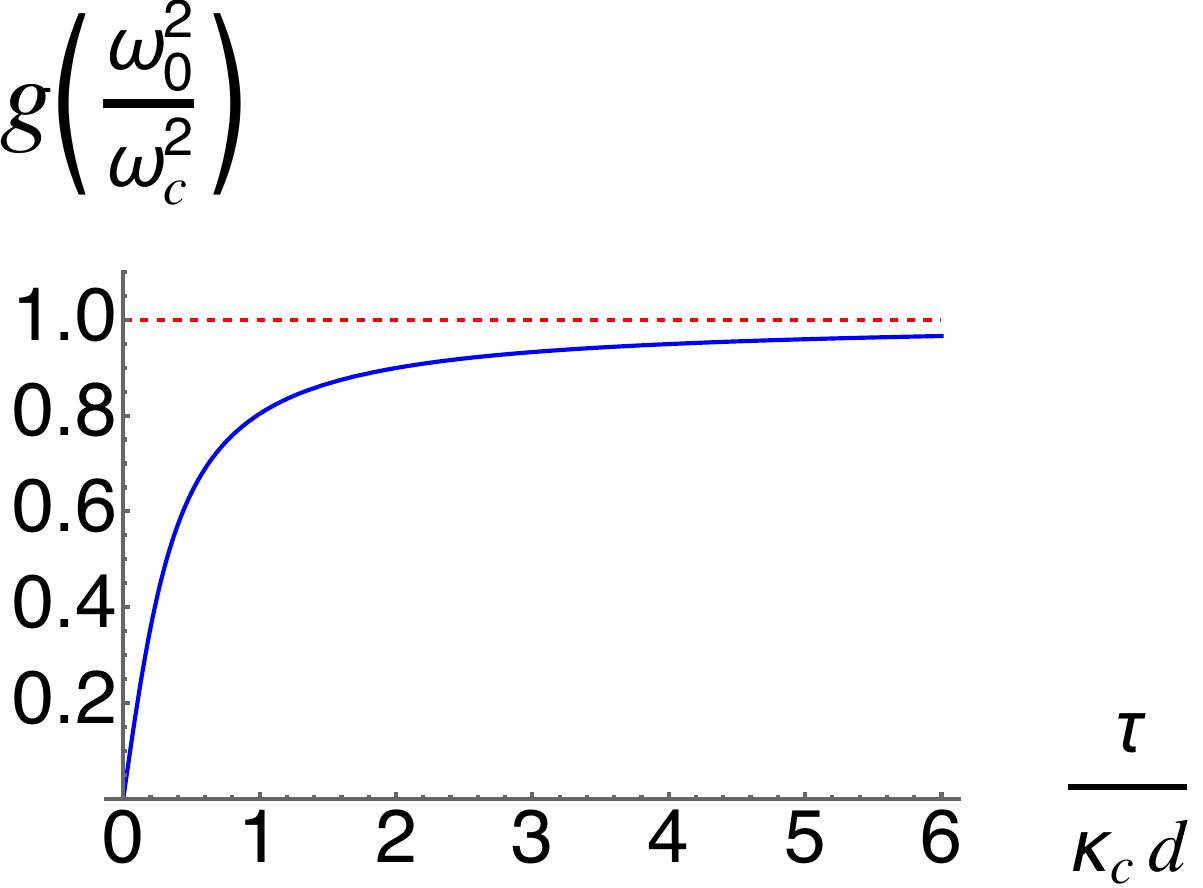}
\caption{\label{fig:gvsks} Coupling strength $g$ versus $\frac{\tau}{\kappa_c d}$ for $N,\ell \to\infty$.}
\end{figure}

We note that the Hamiltonian is no longer positive-definite when the lowest Bogoliubov frequency vanishes.  Thus, there is a constraint on the model; namely, using Eq.~\ref{spectrum}, we see that the following condition must be satisfied to prevent an instability
\begin{equation}
\frac{4}{\omega_0} \sum_n {\frac{\gamma_n^2}{\omega_n}}<1.
\label{critical}
\end{equation}
We define the coupling strength $g\equiv \frac{4}{\omega_0} \sum_n {\frac{\gamma_n^2}{\omega_n}}$ and from Eq.~\ref{critical} conclude that there is a critical coupling strength $g_c=1$ above which the model is ill-defined.  Using the form for $\gamma_n$ in Eq.~\ref{gamma}, we obtain an expression for $g$ in the thermodynamic limit
\begin{equation}
g=\frac{2}{\pi}\bigg(\frac{\omega_c}{\omega_0}\bigg)^2 \tan^{-1}\frac{\pi\tau}{\kappa_c d}
\end{equation}
where $d=\ell/N$, the interatomic distance between atoms in the string.  With increasing string tension $\tau$, $g$ asymptotically approaches $g=\big(\frac{\omega_c}{\omega_0}\big)^2$, a quantity bounded by 1 (see Fig.~\ref{fig:gvsks}). Thus, the model is stable over the range of physical parameters. (As a practical matter, before the tension $\tau$ becomes comparable to interatomic forces in the string, it is likely that the string would break.)

\section{Multimode squeezed vacuum}
Eigenstates of the system can be labeled by the set of Bogoliubov excitation numbers for the $N+1$ modes $|\{n_\alpha\}\rangle$ with corresponding energies $E(\{n_\alpha\})=\sum_\alpha n_\alpha\Omega_\alpha+\frac{1}{2}\sum_\alpha(\Omega_\alpha-\omega_\alpha)$.  The ground state of the coupled system $|\{0\}\rangle$ can be constructed from the uncoupled ground state $|\rangle_0$ with the squeeze operator $S({\xi})=\exp\left(-\frac{1}{2}\sum_{\alpha\beta} \xi_{\alpha\beta} a^\dagger_\alpha a^\dagger_\beta\right)$:
\begin{equation}
|\{0\}\rangle={\cal N}\ S({\xi})|\rangle_0.
\label{svs}
\end{equation}
$\xi$ is the (matrix) squeeze parameter and ${\cal N}$ is a normalization factor.  (Using an identity due to Schwinger \cite{schwinger}, we obtain the normalization constant ${\cal N}= \frac{1}{\sqrt{\det \textsf{M}}}$.)

We verify this by operating on Eq. \ref{svs} with $b_\alpha$ and using the identity $S(-{\xi})a_\alpha S({\xi})=a_\alpha-\sum_{\beta}\xi_{\alpha\beta} a_\beta^\dagger$. We find that Eq. \ref{svs} is satisfied, provided the squeeze matrix has the value
\begin{equation}
\xi=\textsf{M}^{-1} \ \textsf{N}.
\label{squeezemat}
\end{equation}
Hence, the ground state of the model is always a multimode squeezed vacuum state \cite{Ma-Rhodes, coherentstates} with squeeze parameter $\xi$ determined by the Bogoliubov coefficients.

We note that the position uncertainty of the bead in the ground state can be readily evaluated with the Bogoliubov coefficients.  Evaluating the bead variance gives  
\begin{eqnarray}
\langle \{0\}|u_0^2|\{0\} \rangle&=&\frac{\hbar}{2 m \omega_0}\sum_\alpha (M_{\alpha 0}-N_{\alpha 0})^2\\
&=&\frac{\hbar}{2m}\sum_\alpha \frac{1}{\Omega_\alpha(1+4\omega_0\sum_q \frac{\gamma_q^2\omega_q}{(\Omega_\alpha^2-\omega_q^2)^2})}
\end{eqnarray}
The sum can be evaluated analytically using complex contour integration \cite{SM} to reveal that the position uncertainty is reduced with increasing damping $\nu$, a result first obtained by Caldeira and Leggett \cite{leggett} using the fluctuation-dissipation theorem.

\begin{figure}[h]
\includegraphics[width=8cm]{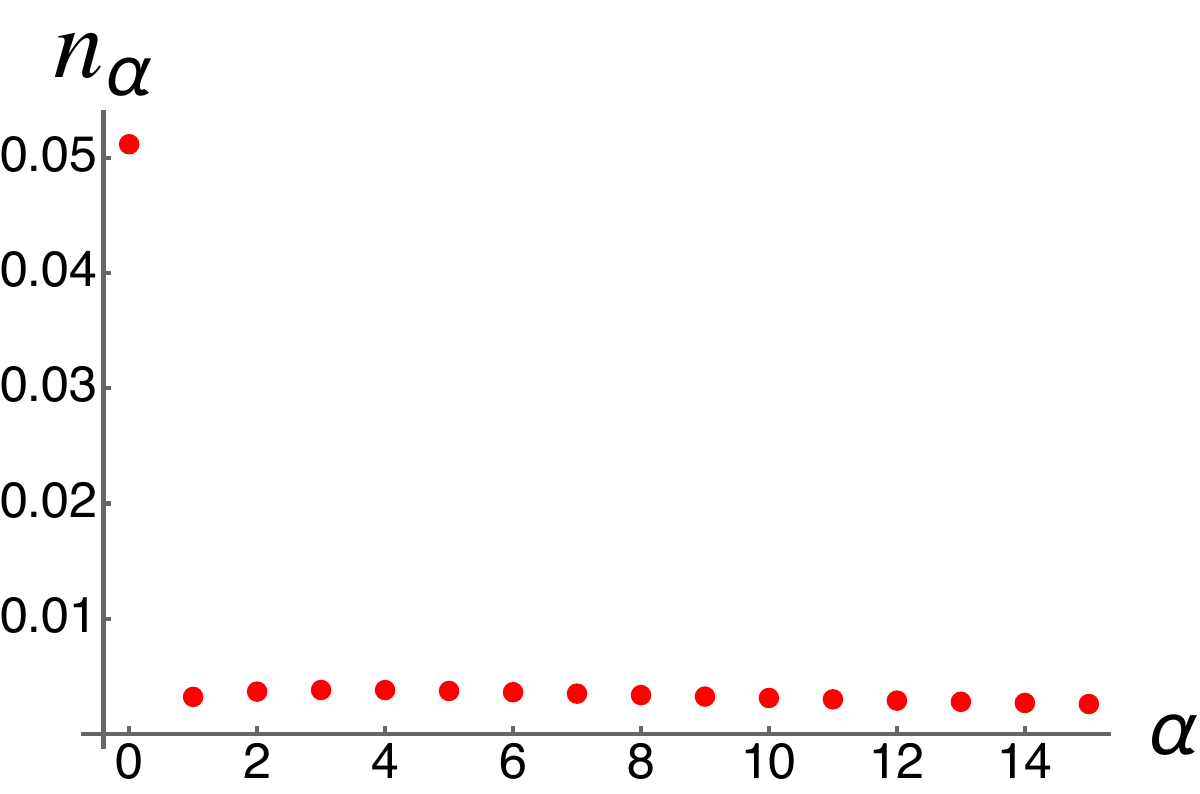}
\caption{\label{fig:nphonons} Distribution of the average number of uncoupled excitations (phonons and vibrons) in the coupled ground state $n_\alpha =\langle \{0\}|a^\dagger_\alpha a_\alpha|\{0\}\rangle$ for parameter values $N=15$, $g=0.7$.}
\end{figure}

To better understand the ground state, the average number of uncoupled excitations (phonons and vibrons) in the mode $\alpha$ contained in the coupled ground state $|\{0\}\rangle$ can also be expressed in terms of Bogoliubov coefficients, with 
\begin{equation}
n_\alpha=\sum_\beta N_{\beta\alpha}^2.
\end{equation}
An example is given in Fig.~\ref{fig:nphonons} for a coupling strength of $g=0.7$. There is a small fraction of a vibron contributed by the bead ($\alpha=0$), with an equal total amount of phonons on the string approximately uniformly distributed across the modes at this coupling strength.

\section{Vibron decay}
\begin{figure}[h]
\includegraphics[width=8cm]{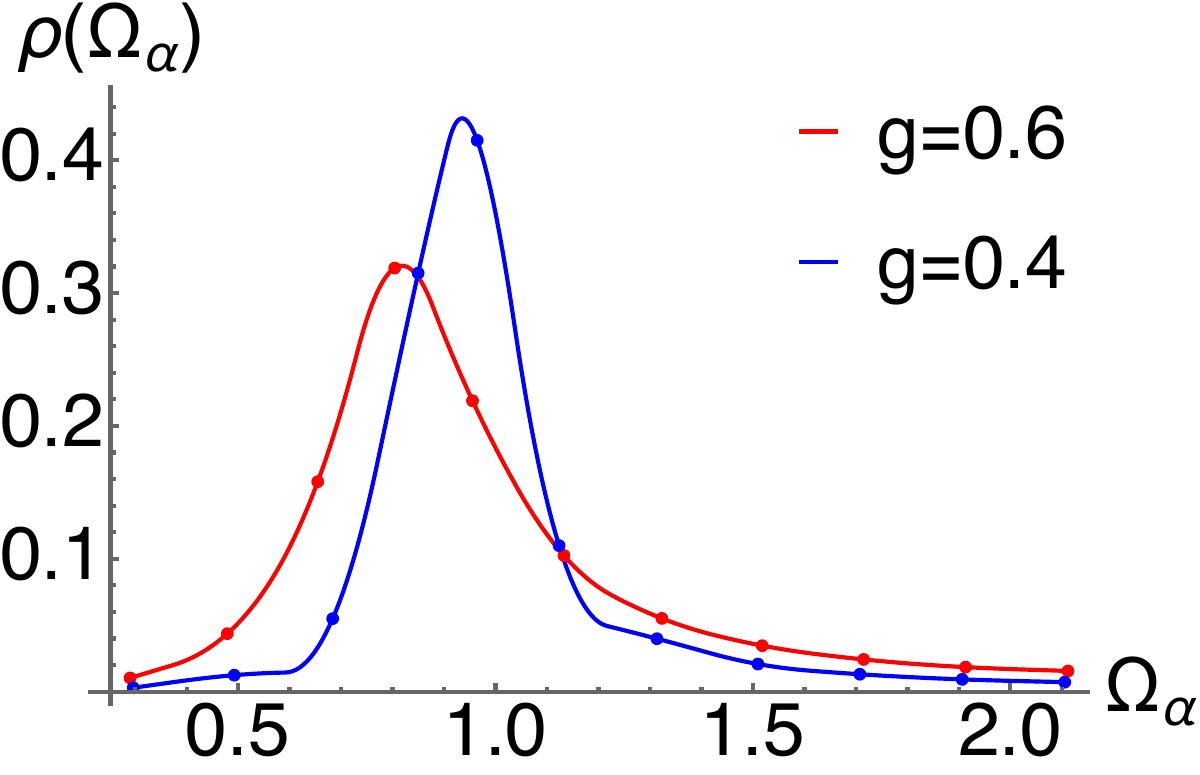}
\caption{\label{fig:probabilty} Spectral distribution function $\rho(\Omega_\alpha)$  versus $\Omega_\alpha/\omega_0$ for $g=0.4$ and $0.6$.  It satisfies the sum rule $\sum_\alpha \rho(\Omega_\alpha)=1$ and its width gives the decay rate $\Gamma$ of $\langle u_0(t)\rangle$.}
\end{figure}

We now consider the dynamics of the vibrational decay of the bead. We start in the ground state $|\{0\}\rangle$, displace the bead by $\delta$ to create the initial state $|\Psi(0)\rangle=\exp(-i p_0 \delta)|\{0\}\rangle$, and compute the expectation of the bead's position at time $t$:
\begin{equation}
\langle u_0(t)\rangle=\langle\Psi(t)|u_0|\Psi(t)\rangle.
\end{equation}
The expectation  can be expressed in terms of Bogoliubov coefficients \cite{SM}
\begin{eqnarray}
\label{spect1}
\langle u_0(t)\rangle&=&\delta\ {\rm Re}\ \sum_\alpha \left(U_{0\alpha}^2-V_{0\alpha}^2\right) \exp(-i \Omega_\alpha t)\\
&=&\delta\ {\rm Re}\ \sum_\alpha {\frac{\exp(-i \Omega_\alpha t)}{1+4\omega_0\sum_n {\frac{\gamma_n^2\omega_n}{\left(\Omega_\alpha^2-\omega_n^2\right)^2}}}}.
\label{spect2}
\end{eqnarray}
We identify the factor $\left(U_{0\alpha}^2-V_{0\alpha}^2\right)$ in the summand of Eq.~\ref{spect1} as the spectral density of the decay: 
\begin{equation}
\rho(\Omega_\alpha)=U_{0\alpha}^2-V_{0\alpha}^2.
\label{rho}
\end{equation}
This spectral density satisfies a sum rule \cite{SM} $\sum_\alpha \rho(\Omega_\alpha)=1$, and the width of this spectral density is the decay rate of the bead displacement \cite{variational} (see Fig.~\ref{fig:probabilty}).

Using contour integration in the complex plane, the sum can be evaluated \cite{SM} and the decay rate $\Gamma$ can be obtained:
\begin{equation}
\Gamma=\frac{\omega_r}{\sqrt{2}}\left(\sqrt{1+\frac{\Gamma_r^4}{\omega_r^4}}-1\right)^\frac{1}{2}
\label{decay}
\end{equation}
where $\omega_r\approx\omega_0$ and $\Gamma_r\approx \sqrt{\nu\omega_0}$.  For the case of light damping where $\Gamma_r\ll\omega_r$, Eq.~\ref{decay} gives $\Gamma\approx \nu$, in agreement with the classical result.

Using Fermi's Golden Rule, we obtain for weak coupling strength the decay rate $\Gamma_{GR}$ for a transition $|1;\{0\}\rangle\to |0;\{1_n\}\rangle$ with the bead losing $\Delta E=\omega_0$ to the string
\begin{eqnarray}
\Gamma_{GR}=&2\pi &\sum_n|\langle 1;\{0\}|H_i| 0;\{1_n\}\rangle |^2 \delta(\omega_n-\omega_0)\nonumber\\
=&2\pi&\int d\omega {\cal D}(\omega) \gamma^2(\omega)\delta(\omega-\omega_0)\nonumber\\
=&2\pi& \frac{\nu\omega_0 c}{\ell\omega_0}\frac{\ell}{\pi c}=2\nu
\end{eqnarray}
 That $\Gamma_{GR}$ is twice the decay rate for the bead displacement is expected, since $\Gamma_{GR}$ is the energy decay rate, while $\Gamma$ is the displacement decay rate.  As energy of the bead varies as the square of the oscillation amplitude, $\Gamma_{GR}=2\Gamma$.

We now turn to the radiation spectrum from the vibrating bead.  The decay of the vibrating bead is accompanied by the emission of bogoliubons.  The probability of the emission of a single bogoliubon of frequency $\Omega_\alpha$ can be expressed in terms of Bogoliubov coefficients: 

\begin{equation}
P_1(\Omega_\alpha)=|\langle \{0\}|b_\alpha a_0^\dagger|\rangle_0|^2=
M^{-2}_{0\alpha}\  (\det{M})^{-1}.
\label{1bogol}
\end{equation}

\begin{figure}[h]
\includegraphics[width=9cm]{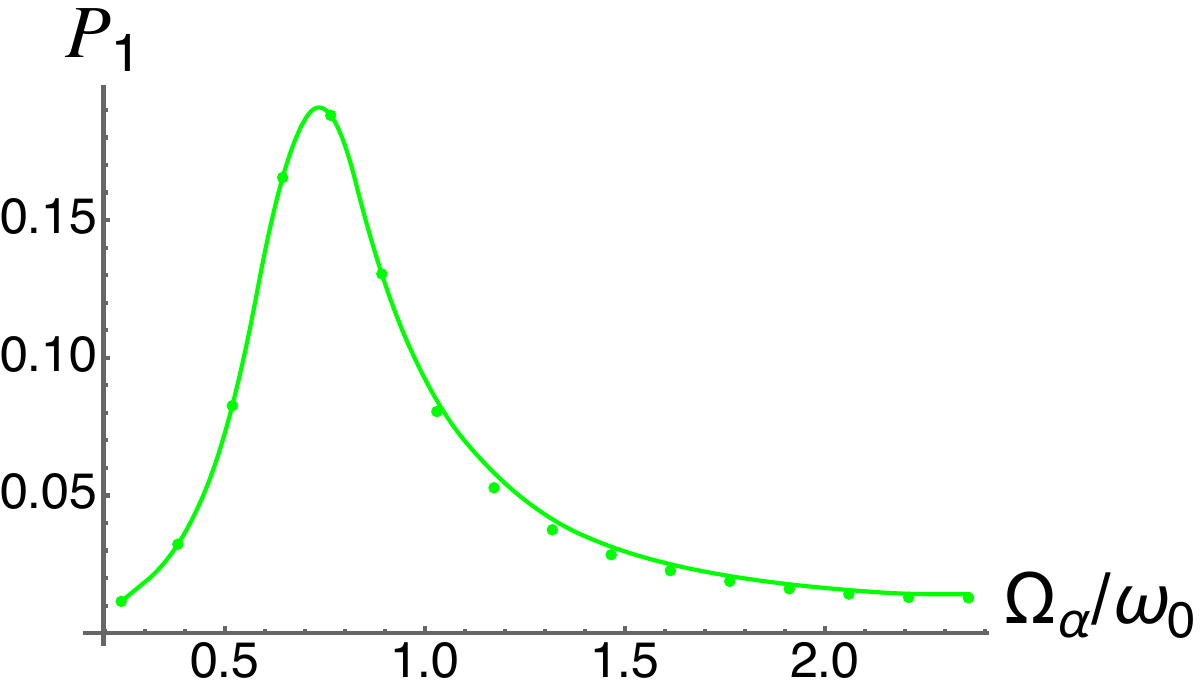}
\caption{\label{fig:1bogol} Spectral probability distribution for single bogoliubon emission $P_1(\Omega_\alpha)$ from the decay of a vibron $|1;\{0\}\rangle$.  Parameter values are $g=0.7$ and $N=15$. The spectrum of single bogoliubons comprise 90.7\% of the total emission.} 
\end{figure}

A plot of the spectral probability distribution for single bogoliubon emission is given in Fig.~\ref{fig:1bogol}.

\section{Summary}
We analyzed the dynamics of a vibrating particle coupled to an environment by extending a generalization of the Lamb model to the quantum regime.  The model provides an exactly solvable example of a damped quantum harmonic oscillator.  These results may apply to a variety of related quantum systems, e.g., a local vibrational mode in a magnetic insulator (vibron-magnon) or coupled to an electromagnetic cavity (vibron-photon).

Our solution explicitly calculates the coefficients of the multimode Bogoliubov transformation that diagonalize the Hamiltonian, and we use these coefficients to describe the properties of the system.  We found that the true ground state of the system is always a multimode squeezed vacuum state where the displacement uncertainty is reduced with increasing damping rate $\nu$.  We then obtained an explicit expression for the vibrational decay rate of the bead and found that it recovered the classical damping rate in the light damping regime.

We examined the acoustic radiation spectrum emitted by the vibrating particle.  We obtained an expression for the probability of single bogoliubon emission in terms of  the Bogoliubov coefficients, and we observed that the spectral emission has a nearly symmetric lineshape about a slightly red-shifted peak frequency.

We thank Dennis Krause for bringing Ref.~\cite{yurke} to our attention. 
This research was supported in part by grant NSF PHY-2309135 to the Kavli Institute for Theoretical Physics (KITP) and grant NASA 80NSSC19M0143.  
\vfil\eject

\bibliography{springlambref}
\vfil\eject

\end{document}


\preprint{arXiv:2503.00562}

\title{Quantum Lamb model: Supplemental Material}

\author{Dennis P. Clougherty}
\author{Nam H. Dinh}
\affiliation{Department of Physics\\
University of Vermont\\
Burlington, VT 05405-0125 USA}

\date{June 17, 2025}

\begin{abstract}
This supplement summarizes some of the relevant properties of symplectic matrices, provides the derivation of the Bogoliubov transformation coefficients used to diagonalize the Hamiltonian for the quantum Lamb model, and gives details on evaluating the sums in Ref.~\cite{main}.
\end{abstract}

\maketitle
 
\newpage
\section{The Model}
\subsection{Classical Formulation}
Consider the Hamiltonian  given by
\begin{equation}
H=H_b+H_s+H_i
\end{equation}
where
\begin{align}
H_b&=\frac{p_0^2}{2m}+\frac{1}{2}m\omega_0^2u_0^2\\
H_s&=\int_0^{\ell}\left[\frac{\Pi^2}{2\sigma}+\frac{\tau}{2}\left(\frac{\partial u_s}{\partial x}\right)^2+\frac{\kappa_c}{2}u_s^2\delta(x)\right] \ \text{d}x\\
H_i&= -\kappa_c u_0 u_s(0,t)
\label{int}
\end{align}
$\omega_0=\sqrt{(\kappa+\kappa_c)/m}$ is the frequency of the bead's oscillations for a fixed string displacement at $x=0$, $u_0(t)$ is the bead's vertical position, $u_s(x,t)$ is the string's displacement field, $p_0$ is the momentum for the bead, $\Pi$ is the string's momentum density, $\sigma$ is the string's lineal mass density, and $\tau$ is the tension in the string.  Such a model recovers the classical equations of motion for the bead and for the string.

 The string displacement $u_s$ is expanded in normal modes of the string, subject to a spring boundary condition at $x=0$
 \begin{equation}
 \tau \frac{\partial u_s}{\partial x}\bigg|_{x=0}=\kappa_c u_s(0,t)
 \end{equation}
 In addition, we apply a fixed boundary condition at $x=\ell$: $u_s(\ell,t)=0$.
 The normal modes of the string $\{w_n(x)\}$ subject to the above boundary conditions are eigenfunctions of a Sturm-Liouville problem.  The eigenfunctions have the form $w_n(x)= A_n \sin k_n(x-\ell)$ where $k_n$ satisfies the transcendental equation
 \begin{equation}
 \tan k_n\ell=-\frac{\tau}{\kappa_c} k_n
 \end{equation}
 and the normalization constant $A_n=\sqrt{\frac{2}{\ell}} \frac{1}{\sqrt{1-\frac{\sin 2 k_n\ell}{2 k_n\ell}}}$.
 We note that $\{w_n(x)\}_{n=1}^\infty$ forms a complete orthonormal set, a result of Sturm-Liouville theory.  Thus, we can expand $\Pi$ and $u_s$ in string normal modes
 \begin{align}
\Pi&=\sum_{n} P_n w_n(x)\\
u_s&=\sum_{n} Q_n w_n(x)
\end{align}

The string and interaction Hamiltonians then become
\begin{align}
H_s&=\sum_n\left(\frac{P_n^2}{2\sigma}+\frac{\sigma\omega_n^2Q_n^2}{2}\right)\\
H_i&= -u_0\sum_n\alpha_nQ_n
\end{align}
where
$\alpha_n\equiv A_n\sin k_n\ell$.

\subsection{Quantum Formulation}
We quantize the Hamiltonian in the standard way to obtain
\begin{equation}
H=\sum_{\alpha}\omega_{\alpha}a_{\alpha}^{\dag}a_{\alpha}-\left(a_0+a_0^{\dag}\right)\sum_{n}\gamma_n\left(a_n+a_n^{\dag}\right)
\end{equation}

The coupling parameters $\gamma_n$ can be expressed in terms of physical quantities of the model 
\begin{equation}
\gamma_n=\omega_x\sqrt{\frac{\nu}{\omega_0}}\sqrt{\frac{k_n\ell}{(k_n\ell)^2+(k_x\ell)^2}}\frac{1}{\sqrt{1-\frac{k_x\ell}{(k_n\ell)^2+(k_x\ell)^2}}}
\label{gamma}
\end{equation}
where $\omega_x=c k_x=\frac{\kappa_c c}{\tau}$ and $\nu=\frac{\tau}{2mc}$.  
(We work with natural units where $\hbar=1$.)

We use a multimode Bogoliubov transformation to bring the Hamiltonian into the form
\begin{equation}
H=\sum_{\alpha}\Omega_{\alpha}b_{\alpha}^{\dag}b_{\alpha}
\end{equation}
with
\begin{equation}
\begin{split}
b_{\alpha}&=\sum_{\beta}\left(M_{\alpha\beta}a_{\beta}+N_{\alpha\beta}a_{\beta}^{\dag}\right)\\
a_{\alpha}&=\sum_{\beta}\left(U_{\alpha\beta}b_{\beta}+V_{\alpha\beta}b_{\beta}^{\dag}\right)
\end{split}
\end{equation}
and $\left[b_{\alpha},b_{\beta}^{\dag}\right]=\delta_{\alpha\beta}$.

\section{Selected properties of Bogoliubov coefficients}
The Bogoliubov coefficients can be grouped to form a $2(N+1)$-dimensional symplectic matrix $\textsf{T}\in \text{Sp}(2(N+1),\mathbb{R})$:
\begin{equation}
\textsf{T}\equiv \begin{pmatrix}
\textsf{M} & \textsf{N}\\
\textsf{N} & \textsf{M}
\end{pmatrix}.
\label{bgt}
\end{equation}
$\textsf{T}$ then satisfies the symplectic condition $\textsf{T} \textsf{J} \textsf{T}^T= \textsf{J}$
where the symplectic form $\textsf{J}$ can be represented as
\begin{equation}
\textsf{J}\equiv \begin{pmatrix}
\ \ \zeroo & \identity\\
-\identity &\zeroo
\end{pmatrix}.
\label{form}
\end{equation}
\subsection{Form of $\textsf{T}^{-1}$}
A number of useful coefficient identities follow from the symplectic structure on this Fock space; for example,  the inverse of the transformation matrix \textsf{T} can be obtained simply from the symplectic condition \cite{arnold} (together with $\textsf{J}^2=-\identity$):
\begin{eqnarray}
\textsf{T}^{-1}&=&- \textsf{J} \textsf{T}^T \textsf{J}\\
&=&\begin{pmatrix}
\textsf{M}^T & -\textsf{N}^T\\
-\textsf{N}^T & \textsf{M}^T
\end{pmatrix}.
\label{ibgt}
\end{eqnarray}
Hence, we conclude that the coefficients of the inverse transformation satisfy $\textsf{U}=\textsf{M}^T$ and $\textsf{V}=-\textsf{N}^T$. From the symplectic condition of \textsf{T}, it is easy to see that $\textsf{T}^{-1}$ is also symplectic, viz., $\textsf{T}^{-1} \textsf{J} \left({\textsf{T}^{-1}}\right)^T=\textsf{J}$.

\subsection{Sum rule and transpose identities:}
Since $\textsf{T}\cdot\textsf{T}^{-1}=\identity$, we conclude that
\begin{enumerate}
\item $\textsf{M}\ \textsf{M}^T-\textsf{N}\ \textsf{N}^T=\identity$ (sum rule)
\item $\textsf{U}\ \textsf{U}^T-\textsf{V}\ \textsf{V}^T=\identity$ (sum rule)
\item $\textsf{U}=\textsf{M}^T$, $\textsf{V}= -\textsf{N}^T$ (transpose rules)
\end{enumerate}
We used these identities in the derivations and to check numerical results.

\subsection{Determinants}
From $\textsf{T}\textsf{J}\textsf{T}^T=\textsf{J}$, one finds that $\det\textsf{J}=(\det\textsf{T})^2\det\textsf{J}$, i.e., 
\begin{equation}
\det\textsf{T}=\pm 1.
\end{equation}
The negative solution however can be ruled out by a simple proof using the Pfaffian \cite{determinants}.

Also, since $\det \textsf{T}=1$, we obtain $\det\left(1-\xi^2\right)=\left(\det \textsf{M}^{-1}\right)^2$ where $\xi=\textsf{M}^{-1} \ \textsf{N}$, the (matrix) squeeze parameter. This result was used to obtain
the normalization factor for the coupled ground state.

\subsection{Symmetries}
From $\textsf{T}\cdot\textsf{T}^{-1}=\identity$, we conclude that
\begin{enumerate}
\item $\textsf{N} \ \textsf{M}^T$ is symmetric
\item $\textsf{U} \ \textsf{V}^T$ is symmetric
\item $\textsf{M}^{-1} \ \textsf{N}$ is also symmetric. This follows from the symmetry of $\textsf{M} \ \textsf{N}^T$.
\end{enumerate}

As $\textsf{M}^{-1} \ \textsf{N}$ is symmetric, we find that 
\begin{equation}
\left(\textsf{M}^T\right)^{-1}=\textsf{M}-\textsf{N}\textsf{M}^{-1}\textsf{N}.
\end{equation}
Thus, $\left(\textsf{M}^T\right)^{-1}$ is a Schur complement of $\textsf{T}$ ($\textsf{T}/\textsf{M}$). This result was used to obtain the spectral probability $P_1$.

\section{Derivation of Bogoliubov Coefficients}
The strategy involves computing commutators with $H$ in two different ways to obtain equations for the unknown coefficients. 

Notice that
\begin{equation}
\begin{split}
[H,b_{\alpha}]&=\sum_{\beta}\Omega_{\beta}\left[b_{\beta}^{\dag}b_{\beta},b_{\alpha}\right]\\
&= -\Omega_{\alpha}b_{\alpha}\\
&= -\Omega_{\alpha}\sum_{\beta}\left(M_{\alpha\beta}a_{\beta}+N_{\alpha\beta}a_{\beta}^{\dag}\right),
\end{split}
\end{equation}
which means that
\begin{equation}\label{way1}
[H,b_{\beta}]= -\Omega_{\beta}\left[M_{\beta0}a_0+N_{\beta0}a_0^{\dag}+\sum_q\left(M_{\beta q}a_q+N_{\beta q}a_q^{\dag}\right)\right].
\end{equation}
A second way of calculating the same commutator gives
\begin{widetext}
\begin{equation}\label{way2}
\begin{split}
&\left[H,M_{\beta0}a_0+N_{\beta0}a_0^{\dag}+\sum_q\left(M_{\beta q}a_q+N_{\beta q}a_q^{\dag}\right)\right]\\
&=\left[\omega_0a_0^{\dag}a_0+\sum_q\omega_qa_q^{\dag}a_q-\left(a_0+a_0^{\dag}\right)\sum_q\gamma_q\left(a_q+a_q^{\dag}\right),M_{\beta0}a_0+N_{\beta0}a_0^{\dag}+\sum_q\left(M_{\beta q}a_q+N_{\beta q}a_q^{\dag}\right)\right]\\
&= -\omega_0a_0M_{\beta0}+\omega_0a_0^{\dag}N_{\beta0}+\sum_q\left(-\omega_qa_qM_{\beta q}+\omega_qa_q^{\dag}N_{\beta q}\right)-N_{\beta0}\sum_q\gamma_q\left(a_q+a_q^{\dag}\right)\\
& \ \ \ \ +M_{\beta0}\sum_q\gamma_q\left(a_q+a_q^{\dag}\right)-\left(a_0+a_0^{\dag}\right)\sum_q\gamma_q(N_{\beta q}-M_{\beta q}).
\end{split}
\end{equation}
\end{widetext}

Equating like-coefficients of creation and annihilation operators in Eqs. \ref{way1} and \ref{way2}, we see that the following system of equations must be satisfied:
\begin{equation}
\begin{split}
-\Omega_{\beta}M_{\beta0}&= -\omega_0M_{\beta0}-\sum_q\gamma_q(N_{\beta q}-M_{\beta q})\\
-\Omega_{\beta}N_{\beta0}&=\omega_0N_{\beta0}-\sum_q\gamma_q(N_{\beta q}-M_{\beta q})\\
-\Omega_{\beta}M_{\beta q}&= -\omega_qM_{\beta q}-N_{\beta0}\gamma_q+M_{\beta0}\gamma_q\\
-\Omega_{\beta}N_{\beta q}&=\omega_qN_{\beta q}-N_{\beta0}\gamma_q+M_{\beta0}\gamma_q
\end{split}
\end{equation}
This gives the following solutions:
\begin{equation}\label{system}
\begin{split}
M_{\beta0}&=\frac{1}{\Omega_{\beta}-\omega_0}\sum_q\gamma_q(N_{\beta q}-M_{\beta q})\\
N_{\beta0}&=\frac{1}{\Omega_{\beta}+\omega_0}\sum_q\gamma_q(N_{\beta q}-M_{\beta q})\\
M_{\beta q}&=\frac{\gamma_q(N_{\beta0}-M_{\beta0})}{\Omega_{\beta}-\omega_q}\\
N_{\beta q}&=\frac{\gamma_q(N_{\beta0}-M_{\beta0})}{\Omega_{\beta}+\omega_q}
\end{split}
\end{equation}
From the sum rule, we have
\begin{equation}
M_{\beta0}^2-N_{\beta0}^2+\sum_q\left(M_{\beta q}^2-N_{\beta q}^2\right)=1.
\end{equation}
Substituting in our expressions for $M_{\beta q}$ and $N_{\beta q}$ found in Eqs. \ref{system}, we find that
\begin{equation}\label{difference of two squares}
\begin{split}
(N_{\beta0}-M_{\beta0})^2\sum_q\gamma_q^2\left[\frac{1}{(\Omega_{\beta}-\omega_q)^2}-\frac{1}{(\Omega_{\beta}+\omega_q)^2}\right]=\\
1+N_{\beta0}^2-M_{\beta0}^2.
\end{split}
\end{equation}
However, from the first two equations in Eqs. \ref{system}, we have that
\begin{equation}\label{Nb0 relation}
N_{\beta0}=\left(\frac{\Omega_{\beta}-\omega_0}{\Omega_{\beta}+\omega_0}\right)M_{\beta0},
\end{equation}
and simplifying the difference of squares in Eq. \ref{difference of two squares}, we find that
\begin{eqnarray}
M_{\beta0}^2\left[1-\left(\frac{\Omega_{\beta}-\omega_0}{\Omega_{\beta}+\omega_0}\right)\right]^2\cdot 4\Omega_{\beta}\sum_q\frac{\gamma_q^2\omega_q}{\left(\Omega_{\beta}^2-\omega_q^2\right)^2}=\nonumber\\
1+\left[\left(\frac{\Omega_{\beta}-\omega_0}{\Omega_{\beta}+\omega_0}\right)^2-1\right]M_{\beta0}^2\nonumber.\\
\end{eqnarray}
Further expanding this out and solving for $M_{\beta0}$, we find that
\begin{equation}
M_{\beta0}=\frac{\Omega_{\beta}+\omega_0}{\sqrt{4\omega_0\Omega_{\beta}}}\frac{1}{\sqrt{1+4\omega_0\sum_q\frac{\gamma_q^2\omega_q}{\left(\Omega_{\beta}^2-\omega_q^2\right)^2}}},
\end{equation}
which means that from Eq.~\ref{Nb0 relation},
\begin{equation}
N_{\beta0}=\frac{\Omega_{\beta}-\omega_0}{\sqrt{4\omega_0\Omega_{\beta}}}\frac{1}{\sqrt{1+4\omega_0\sum_q\frac{\gamma_q^2\omega_q}{\left(\Omega_{\beta}^2-\omega_q^2\right)^2}}}.
\end{equation}
Using the bottom two equations in Eqs. \ref{system}, we can find $M_{\beta q}$ and $N_{\beta q}$:
\begin{equation}
M_{\beta q}= -\frac{2\omega_0\gamma_q}{(\Omega_{\beta}-\omega_q)}\frac{1}{\sqrt{4\omega_0\Omega_{\beta}}}\frac{1}{\sqrt{1+4\omega_0\sum_q\frac{\gamma_q^2\omega_q}{\left(\Omega_{\beta}^2-\omega_q^2\right)^2}}}
\end{equation}
and
\begin{equation}
N_{\beta q}= -\frac{2\omega_0\gamma_q}{(\Omega_{\beta}+\omega_q)}\frac{1}{\sqrt{4\omega_0\Omega_{\beta}}}\frac{1}{\sqrt{1+4\omega_0\sum_q\frac{\gamma_q^2\omega_q}{\left(\Omega_{\beta}^2-\omega_q^2\right)^2}}}.
\end{equation}
From the transpose rules, we obtain $U_{\alpha\beta}=M_{\beta\alpha}$ and $V_{\alpha\beta}= -N_{\beta\alpha}$. 

The table of coefficients is summarized in Table I in Ref.~\cite{main}. The Bogoliubov frequencies are obtained from
\begin{equation}\label{bogofrequencies}
\Omega_{\alpha}^2=\omega_0^2+4\omega_0\sum_k\frac{\gamma_k^2\omega_k}{\Omega_{\alpha}^2-\omega_k^2}.
\end{equation}

\section{Evaluation of Sums}
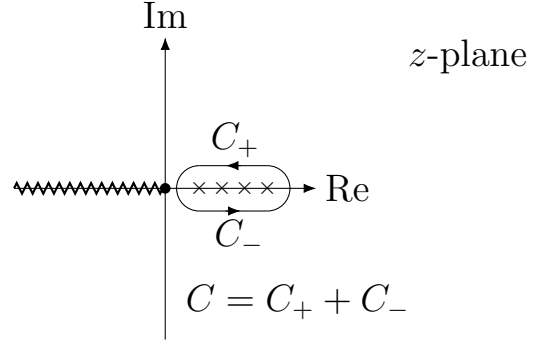
\begin{figure}
\[\begin{tikzpicture}
\draw [-Latex] (-2,0) -- (2,0) node[right,black] {\Large{$\text{Re}$}};
\draw [-Latex] (0,-2) -- (0,2) node[above,black] {\Large{$\text{Im}$}};
\node at (0.45,0) {$\times$};
\node at (0.75,0) {$\times$};
\node at (1.05,0) {$\times$};
\node at (1.35,0) {$\times$};
\node at (4,1.75) {\Large{$z$-plane}};
\draw [thick,decorate,decoration={zigzag,segment length=4,amplitude=2}] (-2,0) -- (0,0);
\draw [-Latex] (0.45,-0.3) -- (1,-0.3) node[midway,below,xshift=2.5mm] {\Large{$C_{-}$}};
\draw (0.8,-0.3) -- (1.35,-0.3) {};
\draw [-Latex] (1.35,0.3) -- (0.8,0.3) node[midway,above,xshift= -1.5mm] {\Large{$C_{+}$}};
\draw (1,0.3) -- (0.45,0.3) {};
\node at (1.75,-1.5) {\Large{$C=C_{+}+C_{-}$}};
\draw (0.45,0.3) arc(90:270:0.3cm and 0.3cm);
\draw [rotate=180] (-1.35,-0.3) arc(270:90:0.3cm and 0.3cm);
\fill [black] (0,0) circle[radius=2pt] {};
\end{tikzpicture}\]
\caption{\label{fig:branch point contour} Closed contour $C$ used to evaluate $I$. The integrand has simple poles on the real axis along with a branch point on the origin. We choose a branch cut along the negative real axis.}
\end{figure}

To calculate the decay rate of the vibrating bead, we start the system in the ground state, $|\{0\}\rangle$, and displace the bead a  distance $\delta$. Therefore, the initial system for this case is
\begin{equation}
|\Psi(0)\rangle=e^{-ip_0\delta}|\{0\}\rangle
\end{equation}
where $e^{-ip_0\delta}$ is the translation operator for the bead; it displaces it by a small amount of $\delta$. We rewrite $p_0$ in terms of the bead's creation and annihilation operators:
\begin{equation}
p_0=i\sqrt{\frac{m\omega_0}{2}}\left(a_0^{\dag}-a_0\right).
\end{equation}
We then calculate the expectation value of $u_0$ over time
\begin{equation}
\langle u_0(t)\rangle=\langle\Psi(t)|u_0|\Psi(t)\rangle.
\end{equation}
We first rewrite $u_0$ in terms of the creation and annihilation operators of the bead:
\begin{equation}
u_0=\frac{1}{\sqrt{2m\omega_0}}\left(a_0+a_0^{\dag}\right).
\end{equation}
This gives us that
\begin{equation}
\langle u_0(t)\rangle=\frac{1}{\sqrt{2m\omega_0}}\left\langle\Psi(t)\left|a_0+a_0^{\dag}\right|\Psi(t)\right\rangle
\end{equation}
with
\begin{equation}
\begin{split}
|\Psi(t)\rangle&=\exp(-iHt)\exp(-ip_0\delta)|\{0\}\rangle\\
\end{split}
\end{equation}
We use the multimode Bogoliubov transformation
\begin{equation}
\begin{split}
b_{\alpha}&=\sum_{\beta}\left(M_{\alpha\beta}a_{\beta}+N_{\alpha\beta}a_{\beta}^{\dag}\right)\\
a_{\alpha}&=\sum_{\beta}\left(U_{\alpha\beta}b_{\beta}+V_{\alpha\beta}b_{\beta}^{\dag}\right)
\end{split}
\end{equation}
together with the transpose rules $U_{\alpha\beta}=M_{\beta\alpha}$ and $V_{\alpha\beta}= -N_{\beta\alpha}$ and the BCH identity to obtain
\begin{equation}\label{sum expectation u0}
\begin{split}
\langle u_0(t)\rangle&=\delta\cdot\text{Re} \sum_{\beta}\left(U_{0\beta}^2-V_{0\beta}^2\right)e^{-i\Omega_{\beta}t}\\
&=\delta\cdot\text{Re} \sum_{\beta}\frac{e^{-i\Omega_{\beta}t}}{1+4\omega_0\sum_q\frac{\gamma_q^2\omega_q}{\left(\Omega_{\alpha}^2-\omega_q^2\right)^2}}.
\end{split}
\end{equation}
To evaluate the sum in Eq.~\ref{sum expectation u0}, we rewrite it as a contour integral in the complex plane. We consider the integral
\begin{equation}\label{initial contour integral}
I=\frac{1}{2\pi i}\oint_C\frac{e^{-it\sqrt{z}}}{z-\omega_0^2-4\omega_0\sum_k\frac{\gamma_k^2\omega_k}{z-\omega_k^2}} \ \text{d}z
\end{equation}
where $C$ is the closed contour pictured in Fig.~\ref{fig:branch point contour}.

We show that $I$ recovers the desired sum. From the residue theorem,
\begin{equation}
I=\sum_{\alpha}\text{Res}(f(z);z=z_\alpha).
\end{equation}
where $f(z)$ is the integrand in Eq.~\ref{initial contour integral}.

The poles of $f(z)$ are located at
\begin{equation}
z_\alpha-\omega_0^2-4\omega_0\sum_k\frac{\gamma_k^2\omega_k}{z_\alpha-\omega_k^2}=0.
\end{equation}
Comparing this to Eq.~\ref{bogofrequencies}, we conclude that the poles are located at $z_{\alpha}=\Omega_{\alpha}^2$, $\alpha=0,1,\ldots,N$.

Hence,
\begin{equation}
I=\sum_{\alpha}\frac{e^{-i\Omega_{\alpha}t}}{1+4\omega_0\sum_k\frac{\gamma_k^2\omega_k}{(\Omega_{\alpha}^2-\omega_k^2)^2}}
\label{sumI}
\end{equation}
as required.

We write the denominator of the integrand $f(z)$ as $z-F(z)$ and consider $N$ to be sufficiently large that we can treat the sum in the quasicontinuum approximation. We replace the sum by an integral in $\omega$:
\begin{equation}\label{f(z) 1}
F(z)=\omega_0^2+4\omega_0\mathcal{D}\int_0^{\omega_d}\frac{\gamma^2(\omega)\omega}{z-\omega^2} \ \text{d}\omega.
\end{equation}
Here, the vibrational density of states of the string is
$\mathcal{D}=\frac{\ell}{\pi c}$,
and the high frequency cutoff is
$\omega_d=\left(\frac{N}{\ell}\right)\pi c$.

$F(z)$ is ill-defined on the positive real axis, but it is a respectable function for $z=x+i\delta$, where $\delta$ is real, small, and positive. We define 
\begin{equation}\label{f plus minus (z)}
F_{\pm}(x)\equiv\omega_0^2+4\omega_0\mathcal{D}\int_0^{\omega_d}\frac{\gamma^2(\omega)\omega}{x-\omega^2\pm i\delta} \ \text{d}\omega \ \left(\delta\rightarrow0^{+}\right).
\end{equation}
We can rewrite $I$ as
\begin{equation}\label{I plus minus}
\begin{split}
I&=\frac{1}{2\pi i}\int_0^{\infty}\frac{e^{-it\sqrt{x}}}{x-F_{-}(x)} \ \text{d}x-\frac{1}{2\pi i}\int_0^{\infty}\frac{e^{-it\sqrt{x}}}{x-F_{+}(x)} \ \text{d}x\\
&=\frac{1}{2\pi i}\int_0^{\infty}e^{-it\sqrt{x}}\left[\frac{1}{x-F_{-}(x)}-\frac{1}{x-F_{+}(x)}\right] \ \text{d}x.
\end{split}
\end{equation}
We evaluate $F_{\pm}(x)$ using Plemelj's identity:
\begin{equation}
\begin{split}
F_{\pm}(x)&=\omega_0^2+4\omega_0\mathcal{D}\fint_0^{\omega_d}\frac{\gamma^2(\omega)\omega}{x-\omega^2} \ \text{d}\omega\\
& \ \ \ \mp4\pi i\omega_0\mathcal{D}\int_0^{\omega_d}\gamma^2(\omega)\omega\delta(\omega^2-x) \ \text{d}\omega\\
&=g(x)\mp ih(x)
\end{split}
\end{equation}
where $\fint$ denotes the Cauchy principal value, and
\begin{align}
g(x)&\equiv\omega_0^2+4\omega_0\mathcal{D}\fint_0^{\omega_d}\frac{\gamma^2(\omega)\omega}{x-\omega^2} \ \text{d}\omega \\
h(x)&\equiv2\pi\omega_0\mathcal{D}\gamma^2\left(\sqrt{x}\right)
\end{align}
are both real-valued functions of $x$. Substituting into Eq.~\ref{I plus minus} gives
\begin{equation}
\begin{split}
    \label{I with g and h}
I=\frac{1}{2\pi i}\int_0^{\infty}e^{-it\sqrt{x}}&\bigg[\frac{1}{x-g(x)-ih(x)}\\
& \ \ -\frac{1}{x-g(x)+ih(x)}\bigg]\ \text{d}x.
\end{split}
\end{equation}
We define $x_r$ such that $x_r=g(x_r)$, and we expand the denominator about $x_r$:
\begin{equation}
\begin{split}
x-g(x)\pm ih(x)&\cong(x-x_r)+[x_r-g(x)]\pm ih(x_r)\\
&\cong(x-x_r)+[x_r-g(x_r)-(x-x_r)g'(x_r)]\\
& \ \ \ \ \pm ih(x_r)\\
&=(x-x_r)[1-g'(x_r)]\pm ih(x_r).
\end{split}
\end{equation}
Substituting into Eq.~\ref{I with g and h} gives
\begin{equation}
\label{approxI}
\begin{split}
I&\approx\frac{1}{2\pi i}\int_0^{\infty}e^{-it\sqrt{x}}\cdot\frac{2ih(x_r)}{[1-g'(x_r)]^2(x-x_r)^2+h^2(x_r)} \ \text{d}x\\
&=\frac{1}{1-g'(x_r)}\cdot\frac{1}{\pi}\int_0^{\infty}e^{-it\sqrt{x}}\cdot\frac{h_r}{(x-x_r)^2+h_r^2} \ \text{d}x\\
\end{split}
\end{equation}
where
\begin{equation}
h_r\equiv\frac{h(x_r)}{1-g'(x_r)}.
\end{equation}
Let $x=\omega^2$. Since the integrand is only appreciable near $\omega_r$, we approximate $I$ by
\begin{equation}
I\cong\frac{2\omega_r\Gamma_r^2}{\pi[1-g'(\sqrt{\omega_r})]}\int_{0}^{\infty}\frac{e^{-i\omega t}}{(\omega^2-\omega_r^2)^2+\Gamma_r^4} \ \text{d}\omega
\end{equation}
where $\Gamma_r^2\equiv h_r$.

From Eq.~\ref{sum expectation u0}, we are only interested in Re $I$:  
\begin{eqnarray}
\text{Re}\ I&\cong&\frac{2\omega_r\Gamma_r^2}{\pi[1-g'(\sqrt{\omega_r})]}\int_{0}^{\infty}\frac{\cos{\omega t}}{(\omega^2-\omega_r^2)^2+\Gamma_r^4} \ \text{d}\omega\nonumber\\
    &=&\bar\Gamma^2\int_{-\infty}^{+\infty}\frac{\cos{\omega t}}{(\omega^2-\omega_r^2)^2+\Gamma_r^4} \ \text{d}\omega\nonumber
\end{eqnarray}
where $\bar\Gamma^2\equiv\frac{\omega_r\Gamma_r^2}{\pi[1-g'(\sqrt{\omega_r})]}$.

\begin{figure}
\[\begin{tikzpicture}
\draw [-Latex] (-3,0) -- (3,0) node[right,black] {\Large{$\text{Re}$}};
\draw [-Latex] (0,-2.5) -- (0,2.5) node[above,black] {\Large{$\text{Im}$}};
\node at (-1.4,0.5) {$\times$};
\node at (-1.25,0.85) {\Large{$z_2$}};
\node at (1.4,0.5) {$\times$};
\node at (1.25,0.85) {\Large{$z_1$}};
\node at (-1.4,-0.5) {$\times$};
\node at (-1.25,-0.85) {\Large{$z_3$}};
\node at (1.4,-0.5) {$\times$};
\node at (1.25,-0.85) {\Large{$z_4$}};
\node at (3,1.5) {\Large{$z$-plane}};
\draw [rotate=180] (2,0)  arc(0:180:2cm and 2cm);
\node [rotate=180] at (0,-2) {\arrowIn};
\node at (-.5,-2.375) {\Large{$C_R$}};
\node at (2.75,-2.1) {\Large{$C=C_x+C_R$}};
\node at (-0.5,0.3) {\Large{$C_x$}};
\node [rotate=0] at (-0.1,0) {\arrowIn};
\draw (0,0) -- (1.4,0.5);
\coordinate (A) at (0,0);
\coordinate (B) at (1.25,0.5);
\coordinate (C) at (2,0);
\pic [draw, angle eccentricity=1.25, angle radius=0.7cm] {angle = C--A--B};
\node at (1.2,0.15) {\Large{$\scriptstyle\theta_0/2$}};
\end{tikzpicture}\]
\caption{\label{fig:simple pole contour} Contour $C$ used to evaluate $J$. $C$ consists of $C_x$, the real axis, and $C_R$, a semicircular contour of radius $R$ in the lower half of the complex plane. The integrand has simple poles at $z=z_n$, $n=1,\ldots,4$.}
\end{figure}
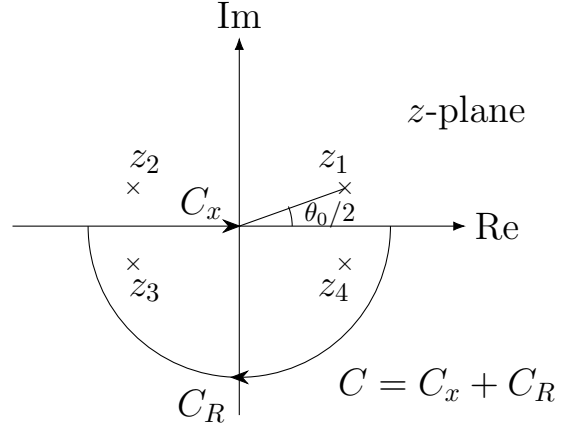

To obtain the time-dependence of $\text{Re}\ I$, we evaluate the integral by considering the complex contour integral
\begin{equation}
J=\bar\Gamma^2\oint_C\frac{e^{-izt}}{(z^2-\omega_r^2)^2+\Gamma_r^4} \ \text{d}z
\end{equation}
 The contour $C$ is shown in Fig.~\ref{fig:simple pole contour}.

From Jordan's lemma,
\begin{equation}
\lim_{R\rightarrow\infty}\int_{C_R}\frac{e^{-izt}}{(z^2-\omega_r^2)^2+\Gamma_r^4} \ \text{d}z\rightarrow0.
\end{equation}
Thus, $\text{Re} \ I=\text{Re} \ J$. We evaluate $J$ using the residue theorem. The integrand of $J$ has 4 simple poles shown in Fig.~\ref{fig:simple pole contour}. The poles are located at
\begin{equation}
\left(z_n^2-\omega_r^2\right)^2+\Gamma_r^4=0,
\end{equation}
so
\begin{equation}
\begin{split}
z_n^2&=\omega_r^2\pm i\Gamma_r^2\\
&=\left(\omega_r^4+\Gamma_r^4\right)^{\frac{1}{2}}e^{\pm i\theta_0},
\end{split}
\end{equation}
where
\begin{equation}
\theta_0=\tan^{-1}\frac{\Gamma_r^2}{\omega_r^2} \ \left(0<\theta_0<\frac{\pi}{2}\right).
\end{equation}
Thus,
\begin{equation}
\begin{split}
J&=2\pi i\bar\Gamma^2\bigg[\text{Res}\left(\frac{e^{-izt}}{(z^2-\omega_r^2)^2+\Gamma_r^4};z=z_3\right)\\
& \ \ \ \ \ \ \ \ \ \ \ \ +\text{Res}\left(\frac{e^{-izt}}{(z^2-\omega_r^2)^2+\Gamma_r^4};z=z_4\right)\bigg].
\end{split}
\end{equation}
We conclude that the damping rate of $\text{Re} \ I$ is given by $|\text{Im} \ z_3| \ (=|\text{Im} \ z_4|)$. Thus,
\begin{equation}
\begin{split}
\Gamma&=|\text{Im} \ z_4|\\
&=\left(\omega_r^4+\Gamma_r^4\right)^{\frac{1}{4}}\sin\frac{\theta_0}{2}\\
&=\frac{\left(\omega_r^4+\Gamma_r^4\right)^{\frac{1}{4}}}{\sqrt{2}}\left(1-\frac{\omega_r^2}{\sqrt{\omega_r^4+\Gamma_r^4}}\right)^{\frac{1}{2}}\\
&=\frac{1}{\sqrt{2}}\left(\sqrt{\omega_r^4+\Gamma_r^4}-\omega_r^2\right)^{\frac{1}{2}}\\
&=\frac{\omega_r}{\sqrt{2}}\left[\sqrt{1+\left(\frac{\Gamma_r}{\omega_r}\right)^4}-1\right]^{\frac{1}{2}}.
\end{split}
\end{equation}

\begin{figure}[hb]
\includegraphics[width=8.2cm]{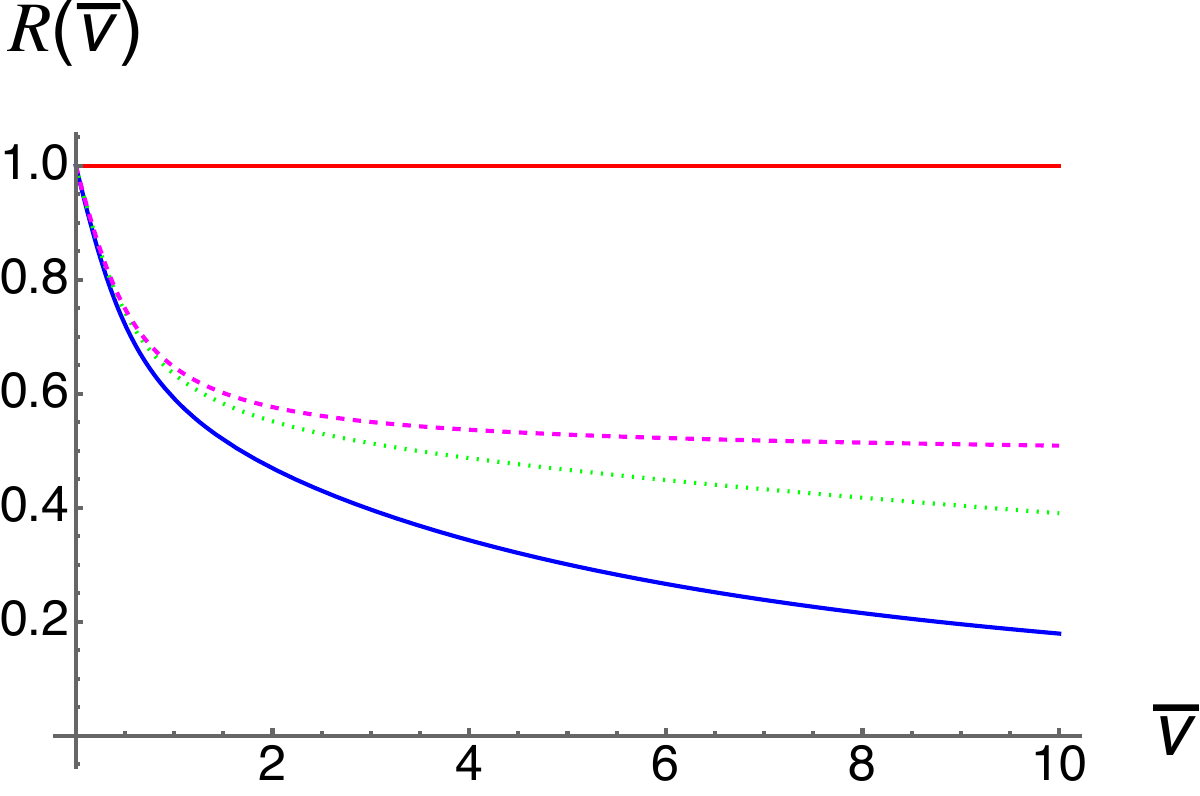}
\caption{\label{fig:uncertainty} Plot of the relative position variance of the bead $R(\bar\nu)$ for $\bar\omega_d=3.5$ (solid blue) ,  $\bar\omega_d=7$ (dotted green), and $\bar\omega_d=30$ (dashed magneta).  The suppression of the bead variance increases with increasing damping.}
\end{figure}

A related sum $K$ needed to evaluate $\langle \{0\}|u_0^2|\{0\} \rangle$ is 
\begin{equation}
K=\sum_\alpha \frac{1}{\Omega_\alpha(1+4\omega_0\sum_q \frac{\gamma_q^2\omega_q}{(\Omega_\alpha^2-\omega_q^2)^2})}
\end{equation}
Following the method used to evaluate the sum $I$, we rewrite $K$ as a complex contour integral
\begin{equation}
K=\frac{1}{2\pi i}\oint_C \frac{2\ dz}{z^2-\omega_0^2-4\omega_0\sum_k \frac{\gamma_k^2 \omega_k}{z^2-\omega_k^2}}
\end{equation}
This integral can be expressed in terms of $F_\pm$ following the logic leading to Eq.~\ref{approxI}
\begin{eqnarray}
K&=&\frac{1}{2\pi i}\int_0^{\omega_d^2} \frac{dx}{\sqrt{x}}\bigg(\frac{1}{x-g(x)-i h(x)}\nonumber\\
&&-\frac{1}{x-g(x)+i h(x)}\bigg)\\
&=&\frac{1}{\pi(1-g'(x_r))}\int_0^{\omega_d^2} \frac{dx}{\sqrt{x}}\frac{h_r}{(x-x_r)^2+h_r^2}\\
&\approx& \frac{1}{\pi\omega_r}\bigg(\arctan\big(\frac{\omega_d^2-\omega_r^2}{2\omega_r\nu}\big)+
\arctan\big(\frac{\omega_r}{2\nu}\big)\bigg)
\end{eqnarray}
The variance of the bead $\langle \{0\}|u_0^2|\{0\} \rangle$ relative to the undamped oscillator is then 
\begin{equation}
R(\bar\nu)=\frac{1}{\pi}\bigg(\arctan\big(\frac{\bar\omega_d^2-1}{2\bar\nu}\big)+
\arctan\big(\frac{1}{2\bar\nu}\big)\bigg)
\end{equation}
where $\bar\nu=\frac{\nu}{\omega_r}$ and $\bar\omega_d=\frac{\omega_d}{\omega_r}$.   The relative variance $R(\bar\nu)$ is plotted in Fig.~\ref{fig:uncertainty}.  The undamped oscillator variance is recovered ($R(0)=1$), and $R<1$ for $\bar\nu>0$ showing the reduction in the variance with increasing damping.

\bibliography{springlambref}